# Effect of Pressure on Electrical and optical Properties of Metal Doped $TiO_2$

Shashi Pandey[1], Alok Shukla[2], Anurag Tripathi[1]

[1]*Department of Electrical Engineering IET Lucknow, Uttar Pradesh 226021, India*

[2]*Department of Physics, Indian Institute of Technology Bombay, Powai, Mumbai 400076, India*

Email: 2512@ietlucknow.ac.in, shukla@iitb.ac.in, anurag.tripathi@ietlucknow.ac.in

**Abstract**

A comparative study of electrical and optical properties of powder and its corresponding pellets has been done on 3d-doped $TiO_2$. $Ti_{1-x}M_xO_2$ (M= Sc, V, Cr, Mn, Fe, Co, Ni, Cu, Zn) powder its corresponding pellet with doping concentration x= 0.05 has also been prepared using solid state route. Optical and electrical measurements have been performed for all prepared samples and interestingly, it is observed that due to having external pressure (i.e. strain) both the properties change significantly. A rigorous theoretical calculation has also been carried out to verify the experimental band gap obtained from optical absorption spectroscopy. In case of pellet sample band gap decreases as compared to the powder sample due to variation of pressure inside the structures. Role of doping has also been investigated both in pellet and powder forms and found that the band gap decreases as the atomic number of dopants increases. A cross-over behavior seen in pellet sample on doping with Ni, Cu and Zn (i.e. band gap increases with increase in atomic number of dopant). Electrical resistivity measurements have been carried out for both pellet and powder samples and it is found that in case of strained samples the value of resistivity is smaller while in case of strain free sample it is quite large. We believe that the present study suggests a novel approach for tuning the electrical and optical properties of semiconducting oxides either from doping or from applied pressure (or strain).

**Introduction**

Recent developments in the field of wide band gap semiconducting oxide materials like $TiO_2$ has stimulated tremendous research efforts [1–3] focused on studying the influence of dopants on their multifunctional properties [4–6]. Despite the intense research in the field of science and technology of semiconductor devices based on GaAs and related group III-V compounds [7], there are still material issues for higher temperatures and pressures that remain to be better understood [8,9]. Effect of pressure and doping in semiconducting oxide-based material has attracted considerable attention of scientific community in recent years because of numerous potential in the field of electrical and opto-electronic industry [1,9–11]. In most of the recent works, the main focus to obtain a sufficient level of photocurrent in the devices[10]. Doping and pressure can also lead to improved optoelectronic and electrical properties in semiconducting oxides based devices[10,12,13].

Driven by growing concerns about environmental and energy issues, interest in semiconductor-based opto-electronic devices has increased considerably over the last few decades [10,11,14,15]. Due to its abundance, non-hazardous nature, and high stability under a variety of conditions, $TiO_2$ is a well-studied material ranging from its synthesis to characterization. Furthermore, numerous experimental and theoretical studies of its physical and chemical properties have already been performed [4], [5], [10]. For this reason, in this manuscript, we do not focus much on the structural analysis, instead try to understand the variation of band gap and electrical properties as a function of its doping with various 3d metals. The electronic band gap of semiconductors tends to decrease with the increasing external pressure[16]. This behavior can be better understood by realizing that overlap between the neighboring atomic orbitals increases with decreasing interatomic distance, leading to enhanced conductivity and reduced band gap. Therefore, direct modulation of the interatomic distance by applying high compressive/tensile stress provides a pathway to tuning of the bandgap[9,17].

The purpose of this article is to study the variation in electrical and optical properties $TiO_2$ as a function of the type of dopant, doping concentration, and external pressure. For the purpose we prepare powder and its pellets of $Ti_{1-x}M_xO_2$ to experimentally study the variation of electrical and optical properties. Additionally, we also theoretically support our experimental work by means of first-principles density-functional theory calculations.

## Experimental and Computational Details

We prepared $Ti_{1-x}M_xO_2$ pellets by solid state route by mixing the dopant (M = Sc, V, Cr, Mn, Fe, Co, Ni, Cu, Zn) and $TiO_2$ powder in required and subsequent to that we add isopropyl alcohol [$CH_3CH_2$ (OH) $CH_3$)] and then grind the sample for one hour. After that we add binder to it in small amount, and grind the sample again for about 15 minutes. Subsequently, we pelletize the powder using 1-inch diset of applied pressure. Next, the pellet is heat treated at 900 $^0C$ for twelve hours finally resulting in a 1-inch target of $Ti_{1-x}M_xO_2$. The band gap of $TiO_2$ and 3d-transition metal doped $TiO_2$ were determined by using UV–VIS spectroscopy with the wavelength range 300-750 nm. Electrical resistivity measurements were done using the four-probe method and the experimental set up consists of probe arrangement, sample, constant current generator, oven power supply and digital panel meter, for measuring the voltage and current. Morphology of pellet and powder samples has been characterized using scanning electron microscopy (SEM) supra Zeiss and Carl Zeiss in plan-views arrangement. All first principles calculations[18] were performed using plane-wave density functional theory (PW-DFT)[19,20] implemented in Quantum Espresso simulation package [21] with generalized gradient approximation (GGA+U)[22–25]. Calculations have been performed on pure $TiO_2$ and 3d- doped $TiO_2$, with supercell of 3x2x2. The k-mesh of size 6 x 6 x 6 in the first Brillouin zone has been used for pure and 5% lattice contracted system. The self-consistent calculations were considered to be converged, when the total energy of the system is stable within $10^{-3}$ mRy, forces per atom declined to less than 0.04 eV/A˚ and the energy convergence up to $5 \times 10^{-5}$ eV[12,26].

## Results and Discussion

The effect of pressure on band gap can be understood in quite simple terms. Pressure changes the lattice parameters and, therefore, the average distance between the ions and the charge carriers. This modifies the potential felt by the charge carriers due to the ions, resulting in the change in the band gap. Comparative morphological study on interfacial strain across grain boundaries has been done on pellet and powder in $TiO_2$. SEM results for pellet and powder samples have been studied and see the grain contribution to produce interfacial strain across phase coexistence. A comparative study of SEM images in Figure-1(a) & 1(b) shows that in pellet sample grain boundaries are closed packed even crystal plane boundaries are well organized with no spacing.

Figure-1(a) & 1(b) also shows pellet samples had comparatively higher density profile with grains than powder sample.

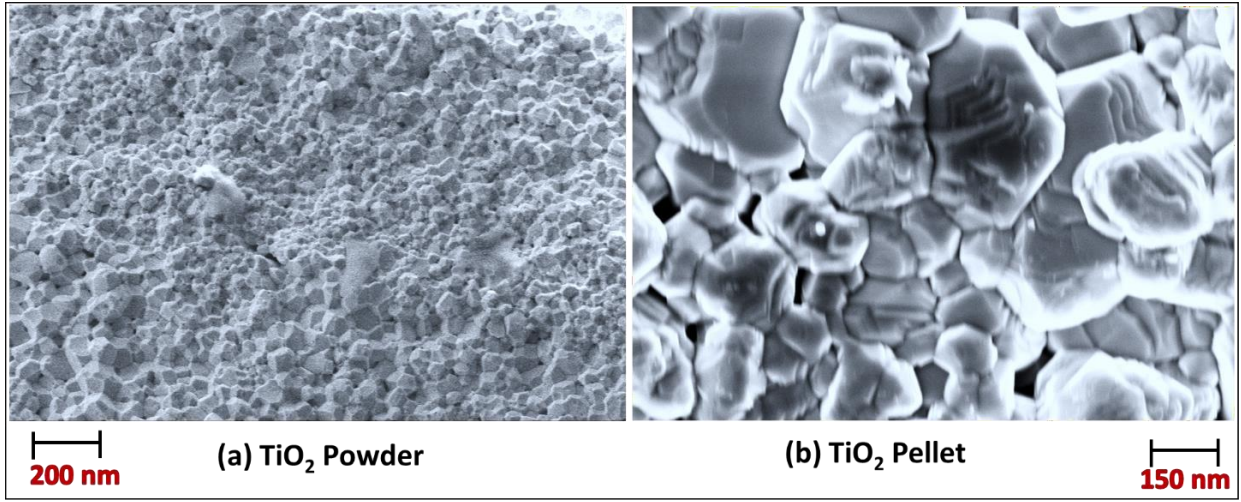

**Figure 1:** *FE-SEM Images of (a) Powder and (b) Pellet.*

In the present study using diffuse reflectance spectroscopy (DRS), we have probed the optical absorption for TiO$_2$ and obtained spectra has been converted into equivalent absorption coefficient using Kubelka–Munk equation[27,28].

$$F(R_\infty) = \frac{(1-R_\infty)^2}{2R_\infty}, \quad \text{(I)}$$

where $F(R_\infty)$ is the Kubelka–Munk function. In order to find the E$_g$, the absorption coefficient is converted in to Tauc equation[29] and plotted in figure 2. The optical gap E$_g$ is determined using the equation-

$$(\alpha h\nu)^n = A(h\nu - E_g), \quad \text{(II)}$$

In equation (II) above we use n=1/2 for the case of pristine TiO$_2$ samples (powder or pellet), because it is an indirect band-gap material. It is observed in figure 2 that the band gap of pellet TiO$_2$ is quite smaller as compared to powder TiO$_2$[30,31]. In case of pellet sample, it is obvious that the amount of strain (i.e. contraction/expansion) is quite large as compared to powder sample leading to the change in optical gap.

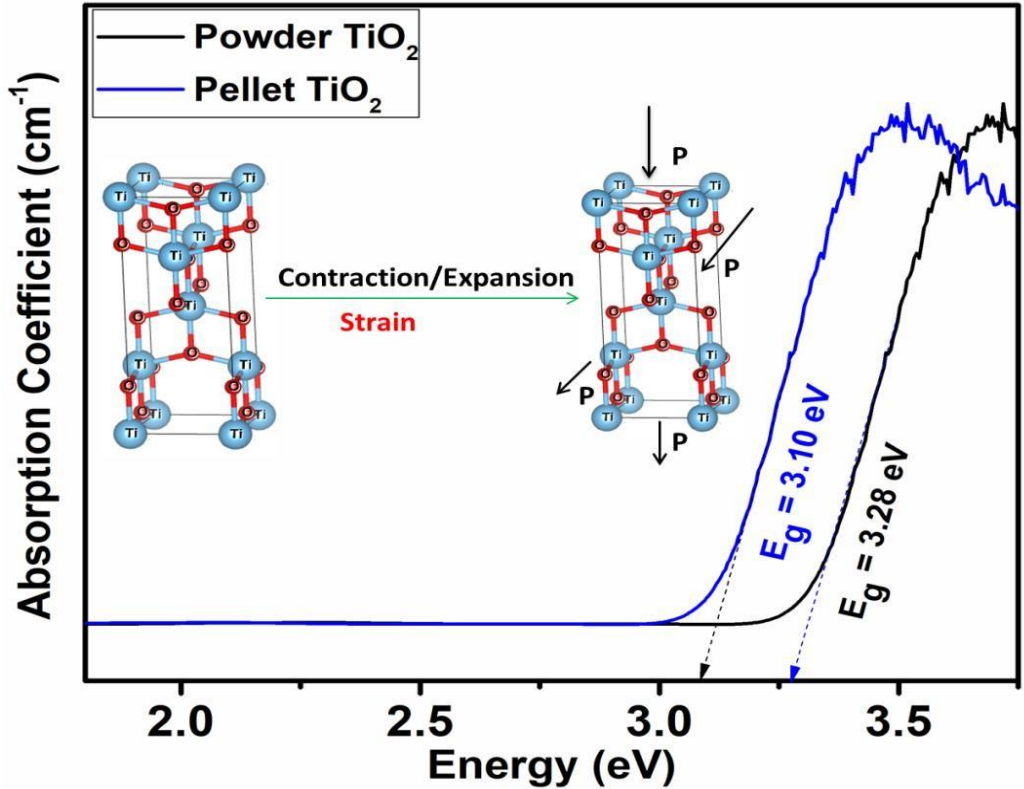

**Figure 2**: *Optical absorption spectra of pellet and powder sample of pure TiO₂.*

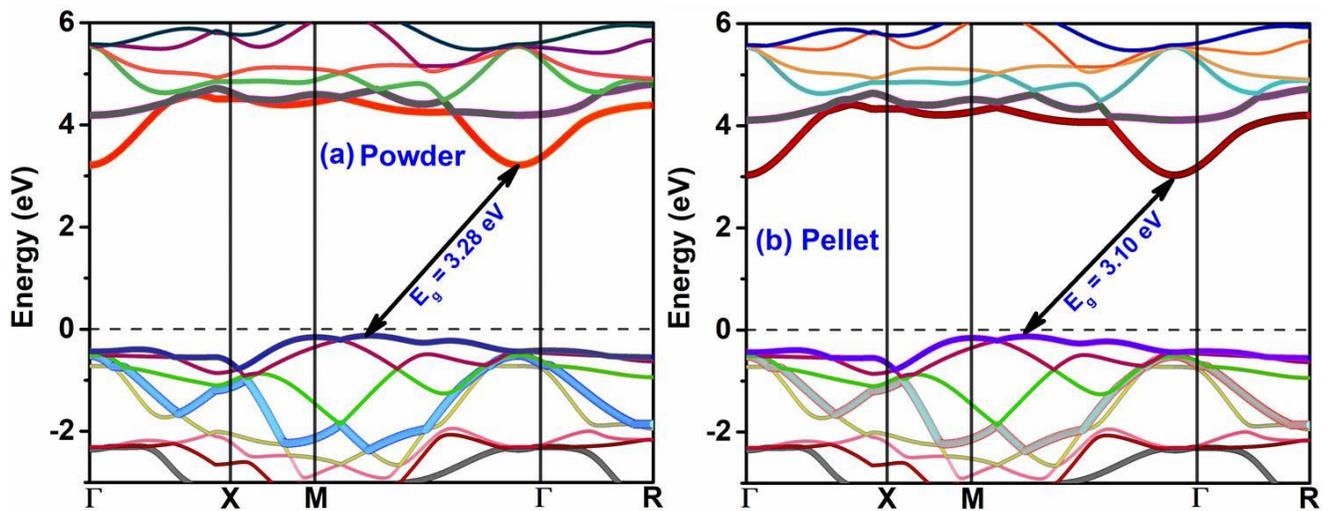

**Figure 3:** Band structure calculations of (a) Powder TiO₂ and (b) Pellet TiO₂.

From Fig. 3 it is obvious that for the pristine TiO₂, the calculated band gap is larger for the powder sample, as compared to the pellet one. Keeping this in view we have also calculated optical band gap for prepared pellet TiO₂ and which is observed smaller band gap with respect to powder samples. From figure 3(b) it is clearly observed that with increase in pressure (strain), the value of band gap decreases with respect to powder sample. Hence, in case of pellet sample internal strain are playing a crucial role in the variation in band gap than powder sample.

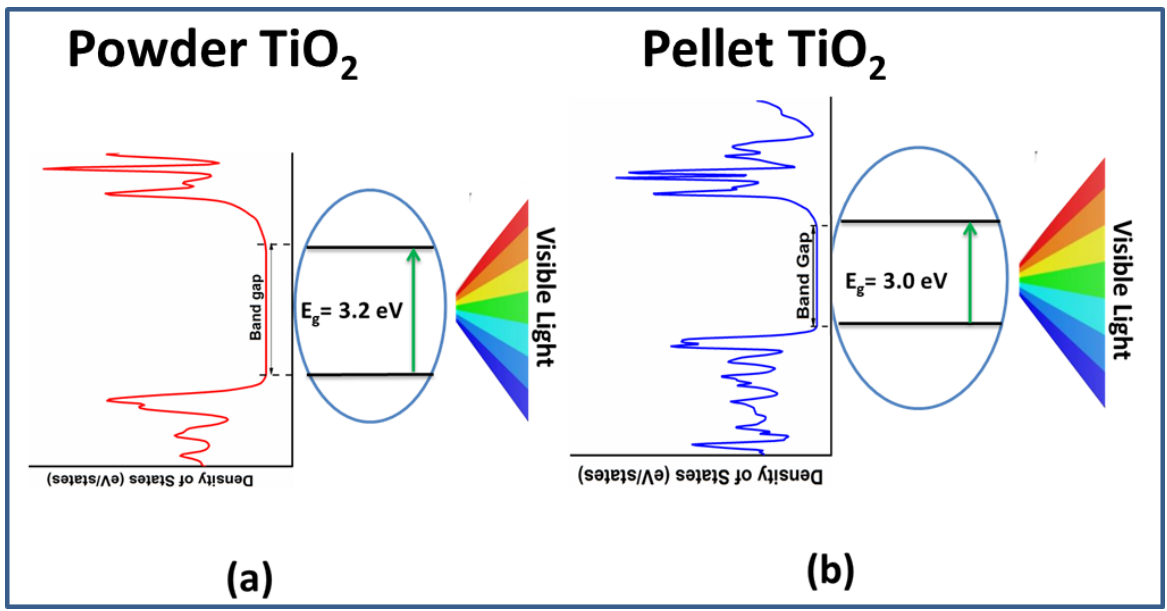

**Figure 4:** *Schematic shows influence of strain (lattice contraction) on electronic band gap of titanium dioxide (a) powder sample (b) pellet sample.*

From figure 4 it is shown that the electronic band gap of pellet $TiO_2$ is comparable smaller than powder samples, which is due to the presence of pressure/strain (i.e. lattice contraction). When UV-VIS light falls on samples, get absorbed by its corresponding wavelength, as a result of this with applied pressure the density of states also shows deviations from 3.2 eV for powder to 3.0 eV for pellet samples. To see the effect of doping with internal strained sample, we have prepared 3d-doped $TiO_2$ elements as a dopant with increase in the atomic number i.e. found from Vanadium (V) to Zinc (Zn) the band gap systematically approaches towards un-doped $TiO_2$. Calculated electronic band gap of pellet and powder of doped $TiO_2$ is shown in the table-1. It is clear from the table that with increase in the atomic number of dopants in $TiO_2$ results decrease in the band gap in both cases, while in case of 3d-doped powder samples have rate of change in band gap is quite large than 3d-dopd pellet samples. In figure 5 shows the variation of band gap of 3d-doped $TiO_2$ for with and without applied pressure. Interestingly, a cross-over behavior has been found in pellet sample in Ni, Cu and Zn doping (i.e. band gap increases with increase in atomic number of dopant).

**Table 1**: *Estimated band gap of 3d-doped TiO₂ pellet and powder*

| Sample | $E_g$ (in eV) pellet | $E_g$ (in eV) Powder |
|---|---|---|
| Pure $TiO_2$ | 3.095 | 3.275 |
| Sc doped $TiO_2$ | 3.10 | 3.28 |
| V doped $TiO_2$ | 3.05 | 3.21 |
| Cr doped $TiO_2$ | 3 | 3.15 |
| Mn doped $TiO_2$ | 2.86 | 3.01 |
| Fe doped $TiO_2$ | 2.7 | 2.86382 |
| Co doped $TiO_2$ | 2.65 | 2.67738 |
| Ni doped $TiO_2$ | 2.572 | 2.54783 |
| Cu doped $TiO_2$ | 2.60157 | 2.48119 |
| Zn doped $TiO_2$ | 2.65429 | 2.44706 |

Figure 6 shows the density of states of Cu doped TiO₂ and it is found that with doping of 3d elements an extra state has been introduced in between the conduction band (CB) and valance band (VB). Formation of new states in between CB and VB shows the signature of defect states in doped titanium dioxides, due to which rehybridization between the 3d-orbital of Ti-atoms and 2p-orbital of O-atoms[10] takes place.

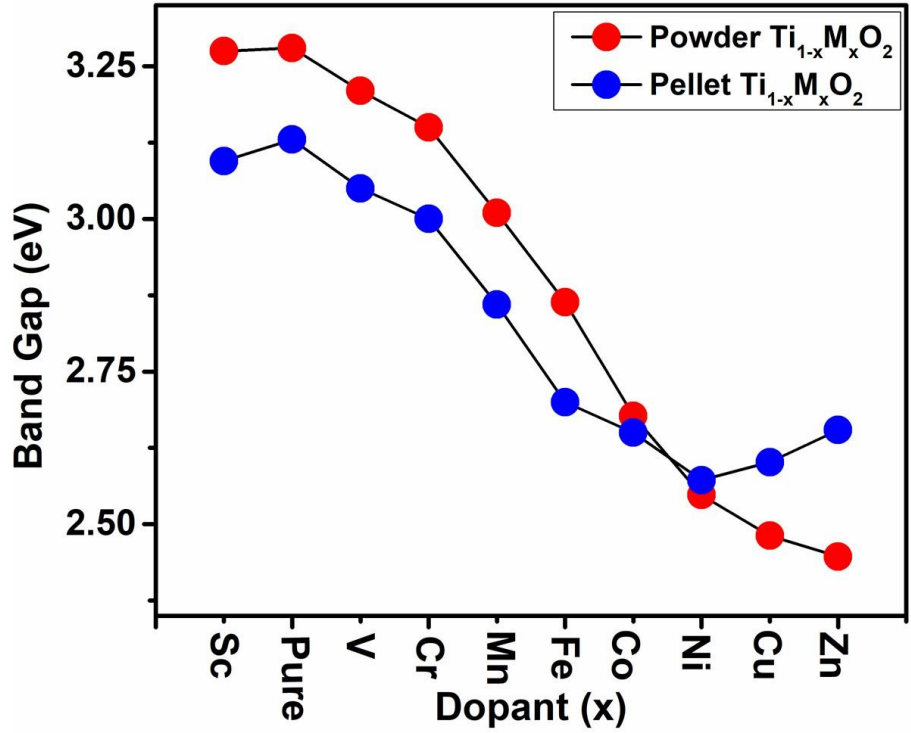

**Figure 5:** *Optical band gap of 3d-doped TiO$_2$ for powder and pellet samples.*

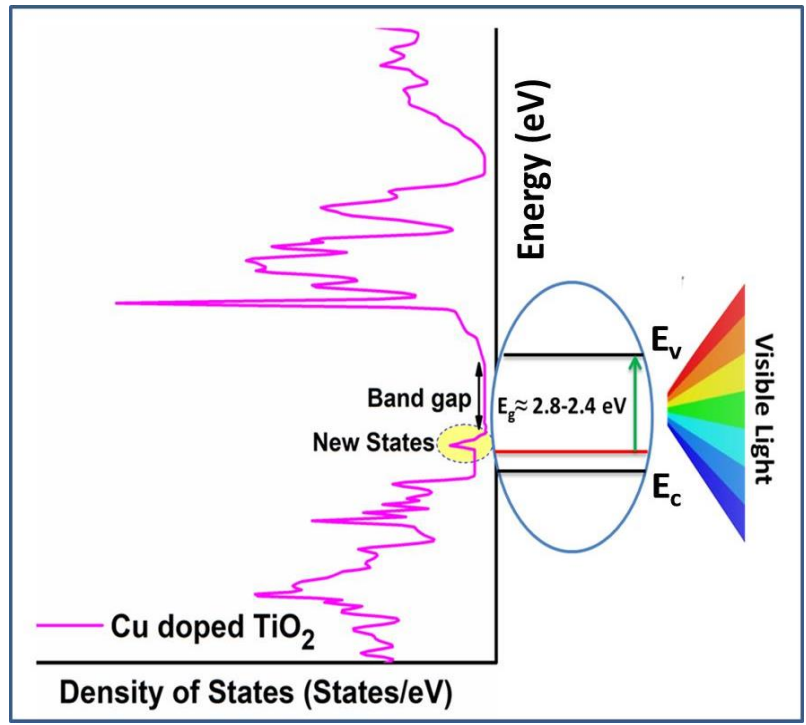

**Figure 6:** *Schematic shows influence of doping on electronic bandgap of doped titanium dioxide.*

Further, to see the effect of strain in electrical properties of doped powder and pellet samples an electrical resistivity measurement has been performed. Electrical resistivity measurement was done from the temperature range 285k-425k for 3d-doped pellet and powder $TiO_2$ samples. From figure 7(a) it is clearly shown that the value of electrical resistivity of 3d-doped powered samples decreases with the application of temperature while, by adding dopant from vanadium to zinc it increases. After identify the effect of doping on electrical resistivity of powder samples, we have also performed the same experiment for pellet samples shown in Figure 7(b). It is very important to note that in case of pellet sample the amount of electrical resistivity observed is ten times higher than powder samples. From figure 7(b) the value of electrical resistivity of 3d-doped pellet samples also decreases with the increase of temperature while, by adding dopant from vanadium to zinc the value of electrical resistivity increases. Hence, doping of 3d-elements plays a vital role for the enhancement of electrical and optical properties.

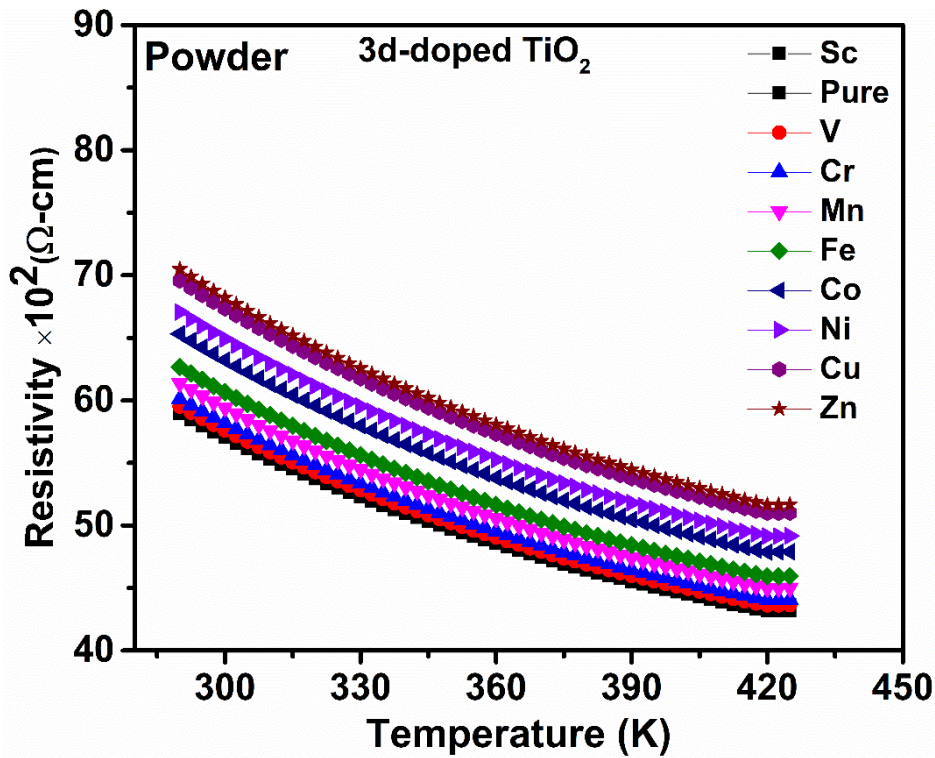

**Figure 7:** *(a) Electrical resistivity of 3d-doped $TiO_2$ for powder samples.*

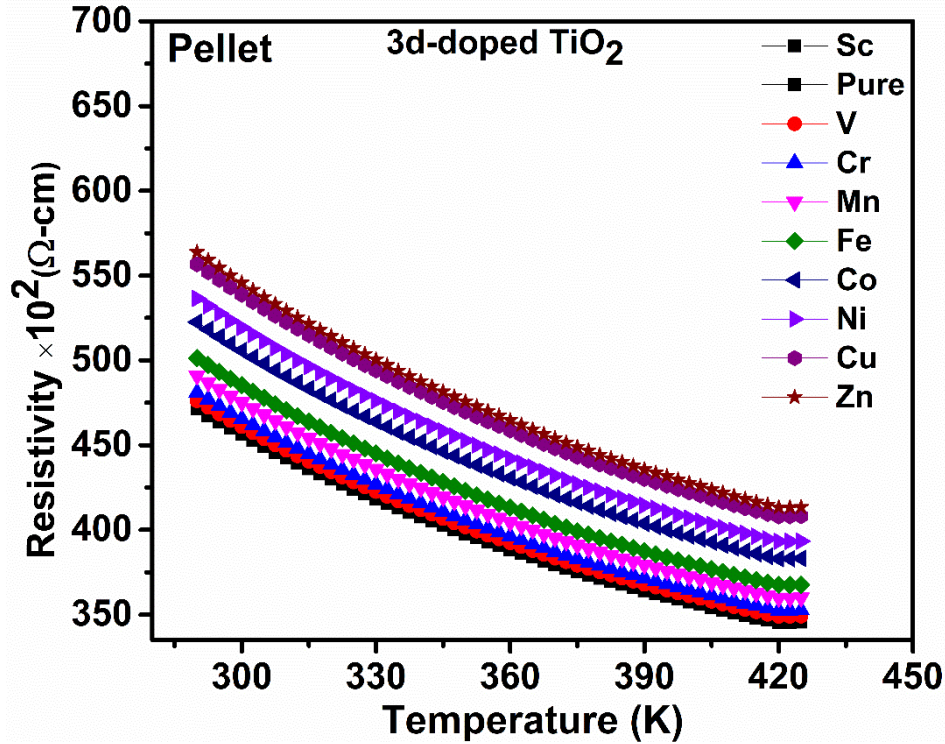

**Figure 7:** *(b) Electrical resistivity of 3d-doped TiO$_2$ for pellet samples.*

**Conclusion**

A combine investigation of electrical and optical properties has been done on 3d-doped-TiO$_2$ powder and its corresponding pellets. Synthesis of powder and pellet samples of doped Ti$_{1-x}$M$_x$O$_2$ (M= Sc, V, Cr, Mn, Fe, Co, Ni, Cu, Zn) have been done. Investigation of optical properties reveals that; pellet samples have smaller optical band gap as compare to powder samples. Electrical resistivity measurements also been performed and it is found that in case of pellet samples the value of electrical resistivity is larger as compare to powder samples. Present study suggests a novel approach for tuning the electrical and optical properties of TiO$_2$ either from doping or from applied pressure (strain).

**Acknowledgments**

One of the authors (SP) acknowledges the Homi Bhabha Research Cum Teaching Fellowship (A.K.T.U.), Lucknow, India for providing financial support Through Teaching Assistantship. Authors sincerely thank Dr. Tejendra Dixit (Assistant Prof. IIIT kancheepuram) for his help during synthesis. One of the authors (SP) likes to thank Dr. Shivendra Pandey (Assistant Prof. NIT Silchar) for optical characterizations and his valuable suggestions in manuscript.

**Data availability statement –**

The raw/processed data required to reproduce these findings cannot be shared at this time as the data also forms part of an ongoing study.

**Compliance with ethical standards:**

Conflict of interest: The authors declare that they do not have any conflict of interest.